\documentclass{ws-procs9x6}
\def\bea{\begin{eqnarray}}
\def\eea{\end{eqnarray}}
\def\bfp{{\bf p}}\def\bfR{{\bf R}}
\def\bfb{{\bf b}}\def\bfk{{\bf k}}
\def\bfq {{\bf q}}
\def\poinc{Poincar\'{e} }
\newcommand{\eq}[1]{Eq.~(\ref{#1})}
\begin{document}

\title{MESON CLOUDS AND NUCLEON ELECTROMAGNETIC FORM FACTORS}

\author{Gerald  A. Miller$^*$}

\address{Department of Physics,
University of Washington\\
Seattle, Washington 98195-1560, USA\\
$^*$E-mail: miller@phys.washington.edu}

\begin{abstract}
In contrast with common non-relativistic lore, the usual Sachs form factors are not
the Fourier transforms of  
charge or magnetization densities.
Instead, the two-dimensional Fourier transform of the 
electromagnetic $F_1$ form factor is the charge 
charge density of partons in the transverse plane. 
 An analysis of the available data for neutron form factors leads to the result 
 that the neutron  charge density 
is negative at the center, and that  the square of the transverse 
charge radius is positive. This  contrasts with many expectations. 
Additionally, the use of measured proton form factors leads to the result
that the proton's central $u$ quark charge density is 
larger than that of the $d$ quark by about 80\%. The proton (neutron) 
charge density has a long range positively (negatively) charged 
component indicative of a pion cloud.
\end{abstract}

\keywords{Generalized Parton Distributions; 
Form Factors, Quark Densities}

\bodymatter

\section{Introduction}\label{sec:intro}
This talk is mainly concerned with the question, ``What do form factors really measure?'',
and secondarily with the question, ``What is the relation between form factors and the
orbital angular momentum of the objects that make  up the neutron or proton?''. The article
\cite{me}, which contains more details,
is the source the present presentation.

A way to focus the discussion is to ask yourself the question, ``What is the charge density 
at the   center of the nucleon?'' The neutron has no net charge, but the charge density 
need not vanish. So we can ask, ``Is the central charge density negative or positive?''.
This talk provides a surprising answer.

There are long-standing existing answers based on models  \cite{cbm,isgurk}.
The neutron can make a spontaneous quantum transition to a state consisting of a proton
and a $\pi^-$ \cite{cbm}. The relatively light pion can spread out over a larger region of space than
the proton. Hence one expects that  the charge density is negative at the edge of the neutron and positive at
the center. The same result is obtained with very different logic from the one-gluon exchange
mechanism 
\cite{isgurk}, which acts repulsively between  two negatively charged d-quarks.
But enough talk about models! Here  we shall be
concerned with model independent information.

\section{Outline}
We shall begin with a discussion of electromagnetic form factors. This will be followed by
a discussion of light cone coordinates and the relevant kinematic subgroup of the \poinc
group. Then the definitions of generalized parton distributions GPDs and a bit of math
lead to the key result \cite{soper1} that the two-dimensional Fourier transform of the electromagnetic form
factor $F_1$ gives the charge density $\rho(b)$ where $b$ is the distance from the transverse 
center of mass, and no information about the longitudinal momentum or position is available. 
Then the data analysis will be discussed, and an attempt at providing 
an interpretation will be made. But really, finding an accurate interpretation is anyone's game.

\section{Definitions}
  The  electromagnetic  form
factors are matrix elements of the current operator,  $J^\mu(x)$, between nucleon states
of different momentum:
\bea
\langle p',\lambda'| J^\mu(0)| p,\lambda\rangle =\bar{u}(p',\lambda')\left(\gamma^\mu F_1(Q^2)+i\frac{\sigma^{\mu\alpha}}
{ 2M}q_\alpha F_2(Q^2)
\right) u(p,\lambda),\eea
where the momentum transfer $q_\alpha=p'_\alpha-p_\alpha$ is taken as space-like, so that 
$Q^2\equiv -q^2>0$, and $M$ is the nucleon mass.
 The nucleon polarization states are chosen to be those of 
definite light-cone helicities $\lambda,\lambda'$ \cite{soperh}.  
The charge (Dirac) form factor is $F_1$, normalized such that $F_1(0)$ is the
nucleon charge, and the magnetic (Pauli) 
form factor is $F_2$, normalized such that $F_2(0)$ is the
anomalous magnetic moment. the Sachs form factors\cite{sachs}
\bea
G_E(Q^2)\equiv F_1(Q^2)-\frac{Q^2}{ 4M^2}F_2(Q^2),\; G_M(Q^2)\equiv F_1(Q^2)+F_2(Q^2),\label{sachsff}\eea
were introduced   to provide an expression for the electron-nucleon cross section
(in the one photon exchange approximation) 
that depends on the quantities $G_E^2$ and $G_M^2$ but not
the product $G_E G_M$.
In  the Breit frame, in which 
$\bfp=-\bfp'$,  $G_E$ is the nucleon helicity flip matrix element of $J^0$. Furthermore, 
 the scattering of neutrons from the electron cloud of atoms  measures
the derivative $-d G_E(Q^2)/dQ^2$ at $Q^2=0$, 
 widely interpreted as six times
the mean-square charge radius of the neutron.  However,
any probability or density interpretation of $G_E$ 
is spoiled by a non-zero  value of $Q^2$, no matter how small \cite{me}.
This is because the initial and final states have different  momentum, and therefore
relativistically have different wave functions. The factorization of relative and
center of mass wave functions that is obtained from the non-relativistic Galilean invariance
is not obtained relativistically. The internal wave function depends upon the total momentum of
the nucleon.
Any attempt to analytically correct for the total momentum  by incorporating
 relativistic corrections in a 
$p^2/m_q^2$ type of expansion  would be doomed, by the presence of the 
very light current quark mass, $m_q$, to  be model-dependent.
That is, at small values of $Q^2$, one finds
\bea G_E^n\sim Q^2 (\int d^3r \left(r^2 |\psi|^2 +\frac{C} { m_q^2}\right),\eea
where the first term represents the traditional effect depending on the square of the
wave function and the unknown 
coefficient $C$ represents the correction due to the total momentum
of the system.

\section{Light cone coordinates}
These useful coordinates involve the use of a ``time'' 
\bea
x^+=(ct +z)/\sqrt{2}=(x^0+x^3)/\sqrt{2}.\eea
The corresponding evolution operator is the not the Hamiltonian, $p^0$, but instead
\bea p^-=(P^0-p^3)/\sqrt{2}.\eea The orthogonal spatial coordinate is 
\bea x^-= (x^0- x^3)/\sqrt{2}.\eea
If  one quantizes at  $x^+=0$, then $x^-=\sqrt{2}z$, and this why $x^-$ is thought of as the
spatial variable. The canonically conjugate momentum is given by 
\bea p^+= (p^0+  p^3)/\sqrt{2}.\eea We note that \bea
p_\mu x^\mu=p^-x^++p^+x^--\bfp\cdot\bfb.\eea
The transverse coordinates perpendicular to the 0 and 3 directions are denoted as
$\bfb$ and $\bfp$.

\section{Relativistic formalism--kinematic subgroup of the \poinc
 group}
The Lorentz transformation defined by a transverse velocity ${\bf v}$  has properties
very similar to that of Galilean transformations. Under these transformations
\bea
k^+\rightarrow k^+\\
\bfk\rightarrow\bfk-k^+{\bf v},\label{tvb}
\eea
and $k^-$ transforms so that $k^2=k^+k^--\bfk^2$ is not changed.
Transverse boosts are like non-relativistic boosts according to Eq.~(\ref{tvb}).

This means that one may use \cite{soper1,mbimpact,diehl2} nucleon 
states that are transversely localized.
  The state with transverse center of mass
$\bfR$ set to 0 is formed by taking a  linear superposition of
states of transverse momentum:
\bea
\left|p^+,{\bf R}= {\bf 0},
\lambda\right\rangle
\equiv {\cal N}\int \frac{d^2{\bf p}}{(2\pi)^2} 
\left|p^+,{\bf p}, \lambda \right\rangle.
\label{eq:loc}
\eea
where $\left|p^+,{\bf p}, \lambda \right\rangle$
are light-cone helicity eigenstates
\cite{soperh} and
${\cal N}$ is a normalization factor.
The relevant range of integration
in \eq{eq:loc} must be restricted to $|\bfp|\ll p^+$ to maintain the interpretation
of a nucleon moving with well-defined longitudinal momentum\cite{mb1}. Thus we use
the infinite momentum frame, for which  
the nucleon may accurately be regarded as a  set of a large number of partons.

\section{The main result}

Using  \eq{eq:loc} sets 
the  transverse 
center of momentum of 
a state of  total very large 
momentum $p^+$  to zero, so that
transverse distance $\bfb$ relative to $\bfR$.
can be  defined. 
Thus we may  define a useful combination of quark-field operators:
\bea
\hat{O}_q(x,{\bf b}) \equiv
\int \frac{dx^-}{4\pi}{q}_+^\dagger
\left(-\frac{x^-}{2},{\bf b} \right) 
q_+\left(\frac{x^-}{2},{\bf b}\right) 
e^{ixp^+x^-},
\label{eq:bperp}
\eea 
where the subscript $+$ denotes the use of only independent quark field operators.
The  
impact parameter dependent PDF is defined \cite{mb1} as the matrix element
of this operator in the state of \eq{eq:loc}:
\bea
q(x,{\bf b}) \equiv 
\left\langle p^+,{\bf R}= {\bf 0},
\lambda\right|
\hat{O}_q(x,{\bf b})
\left|p^+,{\bf R}= {\bf 0},
\lambda\right\rangle. 
\label{eq:def1}
\eea

The use of \eq{eq:loc} in \eq{eq:def1} allows one to show \cite{me} that 
$q(x,{\bf b})$ is the two-dimensional Fourier transform of the GPD $H_q$:
\bea q(x,{\bf b})=\int \frac{d^2q}{ (2\pi)^2}e^{i\;\bfq\cdot\bfb}H_q(\xi=0,x,t=-\bfq^2),\label{ft1}
\eea with $H_q$ appearing because the initial and final helicities are each $\lambda$.

One finds a probability interpretation \cite{soper1} by integrating $q(x,{\bf b})$
over all values of $x$. This  sets the value of $x^-$ to 0, so that 
\bea
\int dx\;q(x,{\bf b}) \equiv 
\left\langle p^+,{\bf R}= {\bf 0},
\lambda\right|q_+^\dagger(0,\bfb)
q_+(0,\bfb)
\left|p^+,{\bf R}= {\bf 0},
\lambda\right\rangle,
\label{eq:def2}
\eea
and a density appears in the matrix elelment.
If one multiplies the above relation   by  the quark charge $e_q$ (in units of $e$),
sums over quark flavors,  uses \eq{eq:loc} with
$\hat{O}_q(x,{\bf b})=e^{-i\hat{\bfp}\cdot\bfb}\hat{O}_q(x,{\bf 0}) e^{i\hat{\bfp}\cdot\bfb}$ 
 along with the sum rule relating the GPD to the form factor,
 the resulting infinite-momentum-frame  IMF parton
charge density in transverse space
is 
\bea
\rho(b)\equiv \sum_q e_q\int dx\;q(x,{\bf b})=\int \frac{d^2q}{ (2\pi)^2} F_1(Q^2=\bfq^2)e^{i\;\bfq\cdot\bfb}.
\label{rhob}\eea

\section{Data analysis and results}

We
exploit \eq{rhob} by  using measured form factors
to determine $\rho(b)$. Recent parameterizations \cite{Bradford:2006yz,Kelly:2004hm,
Arrington:2003qk} of $G_E$ and $G_M$ are very useful  so we express 
$F_1$ in terms of $G_E,G_M$. Then
$\rho(b)$ can be expressed as an integral involving
 known functions:
\bea \rho(b)= \int_0^\infty\; \frac{dQ\;Q\;}{  2 \pi}J_0(Q b)\frac {G_E(Q^2)+\tau G_M(Q^2)}{ 1+\tau},\label{use}\eea
with $\tau={Q^2}/{ 4M^2}$ and $J_0$ a cylindrical Bessel function.
mass.

\begin{figure}
\begin{center}
\psfig{file=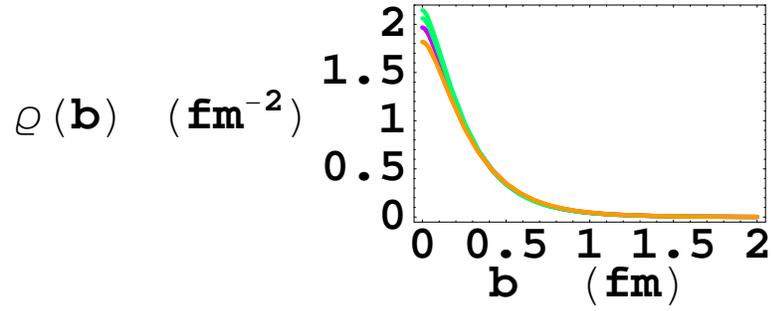,width=4.in}
\end{center}
\caption{The proton charge density $\rho(b) $
using the full spread of the parameters of Kelly's  \cite{Kelly:2004hm} parameterization.
}
\label{fig1}
\end{figure}

\begin{figure}
\begin{center}
\psfig{file=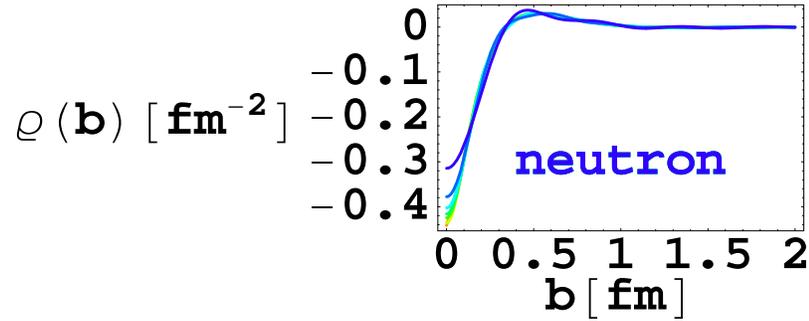,width=4.2in}
\end{center}
\caption{The neutron charge density $\rho(b) $. The upper limit of the integral over
$Q$ in the Fourier transformation \eq{use}, ranges from $Q^2=8 M^2$ to 20 $M^2$, with the
largest (least negative) value at $b=0$ obtained using $8M^2$. 
}
\label{fig2}
\end{figure}

The charge density of the proton is shown in Fig.~\ref{fig1}, and that for the neutron
 in Fig.~\ref{fig2}. The proton
density seems to be well determined, using the entire range of the parameterization
\cite{Kelly:2004hm}, which greatly overestimates the errors, leads to little variation. The 
surprising feature is the negative central value of the neutron
charge density. This results from the negative definite nature of  $F_1$ \cite{me}. 
The neutron density is sensitive to unknown values of $F_1$ at high $Q^2$. Cutting off the
integral appearing in \eq{use} at $Q=2\sqrt{2} M$ leads to big changes, as shown in 
Fig.~\ref{fig2}.

\begin{figure}
\begin{center}
\psfig{file=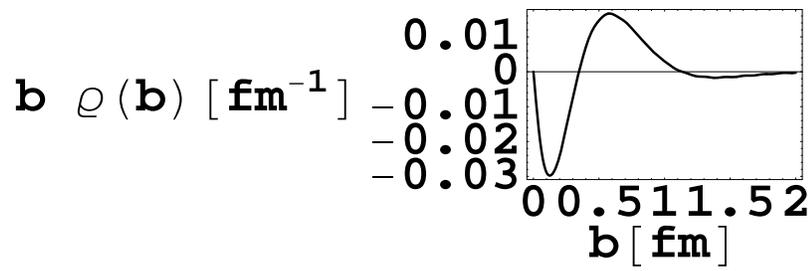,width=4.2in}
\end{center}
\caption{The effective neutron charge density $b\rho(b) $, obtained using
 Kelly's  \cite{Kelly:2004hm} parameterization.} 
\label{fig3}
\end{figure}

The neutron charge density has interesting features, as shown in Fig.~\ref{fig3} which
displays the quantity $b\rho(b)$. It is the integral of this quantity that integrates to 0.
The neutron charge density is negative at the center, positive in the middle, and again
negative  at the outer edge. The medium-ranged positive charge density is sandwiched by
inner and outer regions of negative charge. 

\begin{figure}
\begin{center}
\psfig{file=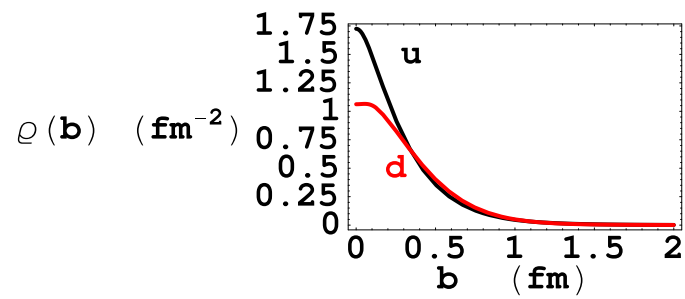,width=4.in}
\end{center}
\caption{The densities for $d$ (red) and $u$ (black) quarks in the proton.}
\label{fig4}
\end{figure}

One can gain information about the individual $u$ and $d$ quark densities by invoking charge symmetry (invariance under a  rotation by $\pi$ about the 
$z$ (charge) axis in isospin space) \cite{mycsb} so that
the $u,d$ densities in the proton are the same as the $d,u$ densities in the neutron.
 We also neglect the effects of $s\bar{s}$ \cite{happex}
or heavier pairs
of quarks. In this case 
$\rho_u(b)=\rho_p(b)-\rho_n(b)/2,\;
\rho_d(b)=\rho_p(b)-2\rho_n(b).$ 
The results, shown in Fig.~\ref{fig4}, and 
are that the central up quark density 
is larger than that of the down quark by about 30\%.

\section{Summary and Interpretation}
Model independent information about parton distributions has been obtained \cite{me}. 
In particular, the central
density of the neutron is negative.
One possible interpretation  is that  quarks with high orbital 
angular momentum do not penetrate the interior, but the negatively charged pion,
with only one unit of orbital angular momentum can both penetrate the interior and
exist at long ranges. If this is the case, the negatively charged 
pions would be suppressed at medium range, allowing
the related density to be positive.

Future  measurements
of neutron electromagnetic form factors could render the present results more precise, or 
 modify
them considerably. Obtaining a qualitative and intuitive understanding of our results presents a challenge
to lattice QCD and to builders of  phenomenological models.

\section*{Acknowledgments}
This work is partially supported by the USDOE.

\end{document}